\documentstyle[sprocl]{article}

\def\be{\begin{equation}}
\def\ee{\end{equation}}
\def\bea{\begin{eqnarray}}
\def\eea{\end{eqnarray}}
\begin{document}

\title{LEFT RIGHT SUSY AND THE FATE OF R-PARITY\footnote{Invited Talk
  presented at PAST-97, ICTP ,
 Trieste October 1997 and
SUSYUNI, TIFR, Mumbai, December 1997, based on work done ion
collaboration with K. Benakli, A. Melfo, A. Ra\v{s}in and G. Senjanovic} }

\author{Charanjit S. Aulakh}

\address{Department of Physics, Panjab University,\\
Chandigarh, INDIA\\E-mail: aulakh\%phys@puniv.chd.nic.in}


\maketitle\abstracts{ A fresh analysis of Left right symmetric
supersymmetric models in the generic case where the scale of
right handed symmetry breaking $M_R >> M_{SUSY}\sim M_W$ is
presented. We conclude that the low energy effective theory for
such models is essentially the MSSM with R parity (and therefore B,L
symmetry) but the spectrum includes heavy conjugate neutrino
supermultiplets that permit a seesaw mechanism and several
characteristic charged supermultiplets over and above those of
the MSSM. }

\section{Introduction}

The Standard Model (SM) has the appealing feature that
( perturbative ) Baryon ( B ) and Lepton ( L ) number conservation follows
automatically from the requirements of renormalizability and
gauge invariance of the lagrangian for the fields observed in
nature (plus the Higgs required for spontaneous symmetry
breaking).  Supersymmetric extensions of the standard model ,
where each of the observed fields of the SM has a partner of
opposite statistics, do not retain this property since the
presence of scalar fields carrying B and L implies that one can
write renormalizable gauge invariant 
interactions that violate B and L. These
terms are described by the superpotential 

\begin{equation}
W_{2} = \mu' L{\bar H} + \lambda_1 u^c d^c d^c +
\lambda_2 Q L d^c + \lambda_3 LLe^c
\label{eq:WRviol}
\end{equation}

These interactions 
 imply ultrarapid proton decay and appreciable amplitudes for
a host of exotic processes unless the couplings $\lambda_i,
\mu'$ are extremely small. The couplings in $W_2$ can be ensured
to vanish if one invokes an ad-hoc $Z_2$ symmetry, the so called
R parity under whose action the superfields pick up \cite{ma92} a phase 
$(-)^{3(B-L)}$  and the supercoordinates $\theta$ 
change sign. Equivalently one
may describe it as a symmetry under which all superpartners of
SM fields are odd while SM fields are even. An
attractive rationale for this symmetry is for it to follow 
automatically from the presence of a gauged $B-L$ symmetry 
\cite{m86,bl89,ma92} at
the supersymmetric level. This would also protect it from the
violation of global symmetries thought to occur due to Planck
scale effects \cite{abs92}. Then the violation of R parity becomes
tied to the spontaneous breaking of the $B-L$ gauge symmetry and can occur
only if a scalar field with an odd value of $B-L$ aquires a vev
\cite{am82,m86,ma92,bl89}. Whether and to what extent this happens has a crucial 
role in determining the low energy phenomenology of the SUSY theory.
In partcular it determines whether the LSP is stable and affects
neutrino masses in interesting and predictable ways \cite{am83}.

The Left Right symmetric model \cite{leftright}
is one of the most appealing ``Next to Minimal'' models since
it provides a ``spontaneous'' rationale for the maximal parity
violation observed in the standard model along with a
 framework to understand the suppression of neutrino
masses. Moreover its $U(1)$ gauge group turns out to be
identical with $U(1)_{B-L}$ so that
while incorporating the standard model it ``takes care'' to
gauge its maximal anomaly free global symmetry ! Thus a
formulation of a minimal left right supersymmetric extension of
the standard model in which R parity is a part of the gauge
symmetry offers the prospect of furnishing a model with
rationales for several problematic features of the standard
model. 

In this talk I shall present recent progress in these directions
made in collaboration with Karim Benakli, Alejandra Melfo,
Andrija Rasin and Goran Senjanovic \cite{abs97,ams97,amrs97} 
and embodied in two workable
models : one with a renormalizable superpotential but an extended
field spectrum that allows decoupling of the scales
($M_R,M_{B-L}$)  at which the
$SU(2)_R$ and $U(1)_{B-L}$ symmetries are broken and another in
which these scales coincide but nonrenormalizable interactions
(upto quartic) are retained in the superpotential. An important
aspect of our work is that we employ a powerful method for the
characterizing the flat directions of supersymmetric vacua by the
holomorphic gauge invariants of chiral supermultiplets left
independent after impostion of the holomorphic constraints 
required by the minimization of the
potential in supersymmetric gauge theories. This method is
useful only for analysis of symmetry breaking at scales where
supersymmetry is a good symmetry. Hence our analysis is
not applicable to the special case 
when the right handed and $B-L$ breaking scales
are of the same order of magnitude \cite{km93,km95}
 as the supersymmetry breaking
scale ($M_S$) but only to the generic case when these scales are much
greater than $M_S$.

The main conclusion of our analysis is that the structure of the
``minimal'' supersymmetric left right vacuaa together with low
energy data imply that unless electromagnetic charge invariance
is violated R parity remains unbroken so that the effective low
energy theory is the MSSM with R parity with certain
characteristic additional supermultiplets in the spectrum.
 The most striking
experimental signature of these theories is the presence of a
number of charged Higgs supermultiplets with masses much less
than $M_R$ including doubly charged particles in the
nonrenormalizable case.

\section{Minimal Supersymmetric Left Right Models}

As mentioned above the gauge group of these models is 
$SU(3)_c \times SU(2)_L \times SU(2)_R \times U(1)_{B-L} $.  The
gauge bosons are accompanied by their corresponding gauginos
while the matter fields of the standard model plus the right
handed neutrino field required by left-right symmetry are
grouped into  three 
 generations of quark and leptonic chiral superfields with the 
following transformation properties:
\begin{eqnarray}
Q=(3,2,1,1/3)&\;\;\;\;\;  & Q_c=(3^*,1,2,-1/3) \nonumber \\ 
L=(1,2,1,-1)& \;\;\;\;\; & L_c=(1,1,2,1) 
\label{qulep}
\end{eqnarray}

The doublet Higgs fields of the standard model are doubled in
number by the requirement of anomaly cancellation in the MSSM
and once again by left right symmetry. Thus they fit into
``bidoublets'' w.r.t the left right gauge group. Realistic
fermion mass matrices require that we introduce at least a pair
of such bidoublets :
\begin{equation}
 \Phi_i=(1,2,2^*,0) \quad (i = 1, 2)
\end{equation}
Where the numbers in brackets give the transformation properties
of the chiral multiplet with respect to the gauge group.

In addition to these fields we must choose Higgs representations
to accomplish the breaking of the $SU(2)_R \times U(1)_{B-L} $
symmetry down to the $U(1)_Y$. The two simplest choices are
pairs of doublets and triplets respectively. The latter choice
allows a see-saw mechanism even at the renormalizable level .
Moreover in the supersymmetric case the doublets  carry
odd $B-L$ so that the $M_{B-L}$ and the scale of R parity
breaking necessarily coincide . This implies that additional ad-hoc
features must be introduced to control the R parity breaking
couplings to lie within the stringent experimental limits. The
triplet case is free of these problems so we shall consider only
this case here. The fields we use are :

\begin{eqnarray}
  \Delta = (1,3,1,2)  , \quad 
  \overline{\Delta}  = (1,3,1,-2) \nonumber \\
\Delta_c = (1,1,3,-2), \quad \overline{\Delta}_c = (1,1,3,2)
\label{minimal}
\end{eqnarray}
The doubling of Higgs triplets here relative to the non
supersymmetric case   again follows from anomaly
cancellation.
Left right symmetry is implemented as a charge conjugation
(convenient when one wishes to embed in a GUT) rather than as
an ( equally feasible and equivalent ) parity transformation :

\begin{eqnarray}
Q              \leftrightarrow     Q_c ,\quad 
L              \leftrightarrow     L_c ,\quad
\Phi_i         \leftrightarrow     {\Phi_i}^T,  \nonumber \\
\Delta         \leftrightarrow  \Delta_c ,\quad
\overline{\Delta}   \leftrightarrow   \overline{\Delta}_c \, .
\end{eqnarray} 

 With this minimal set of Higgs fields , however,
the superpotential for these triplets which are to accomplish
the breaking to $SU(2)_Y$ is merely

\begin{equation}
 W_{LR}= 
       i {\bf f}( L^T \tau_2 \Delta L    
     + L^{cT}\tau_2 \Delta_c L_c ) + 
m_\Delta  ( {\rm  Tr}\, \Delta \overline{\Delta} 
       +   {\rm Tr}\,\Delta_c \overline{\Delta}_c )
   \label{minsuperpot}
\end{equation}

It is easy to see that the vanishing of $F$ terms required by
the minimization of the superpotential now requires that the
vevs of $\Delta, \Delta_c$ vanish while the vevs of
${\bar{\Delta}}$, ${\bar{\Delta}_c}$ are proportional to those of
$fLL$ and $f^*L_cL_c$ respectively. Moreover since squark vevs
break color and charge we assume that they are forbidden by
soft mass terms. The vanishing of $D_{B-L}$ then implies that 
$SU(2)_R$ and $SU(2)_L$ are broken at the same scale. Although
one can evade this problem by introducing a ``parity odd''
singlet \cite{c85} 
it leads to additional difficulties \cite{km95} in achieving a
realistic breaking pattern and is aesthetically displeasing in a
gauge theory to boot. Instead we employ either an additional
pair of $B-L$ neutral triplets 

\begin{equation}
\Omega = (1,3,1,0) , \quad \Omega_c = (1,1,3,0)
\end{equation}
where under Left-Right symmetry $\Omega \leftrightarrow
\Omega_c$,  while maintaining
renormalizability , or allow non-renormalizable terms. 

The potential of a supersymmetric gauge theory has the form

\begin{equation}
V= \sum_{Chiral } |F_i|^2 + \sum_{Gauge } D^2
\end{equation}
and its minimization thus requires that the F and D terms
corresponding to each chiral multiplet and gauge generator
respectively vanish i.e that the vacuum manifold is ``D and F
flat''.  The following powerful theorem \cite{holomorphic} is extremely
useful in characterizing solutions of these equations :

{\bf{Theorem :}} a) The D-flat field space is
coordinatized by the independent Holomorphic gauge invariants
that one can form out of the chiral multiplets.
b) The D and F flat field space is coordinatized by the
Holomorphic invariants left independent after imposition of the
$F=0$ conditions.

As an example : consider a $U(1)$ gauge theory with two chiral
 multiplets $\phi_{\pm}$
with gauge charges $\pm 1$. Then the condition $D=0$ requires only
$|\phi_+|=|\phi_-|$. Since gauge invariance can be used to rotate
away one field phase we are left with a magnitude and a phase i.e one
complex degree of freedom left undetermined. Result a) above predicts
this since the only independent holomorphic gauge invariant in this
case is simply $\phi_+\phi_-$ . Now consider the effects of a
superpotential 
$W=m \phi_+\phi_-$. The F flatness condition now ensures that both
vevs vanish so that the D flat manifold shrinks from the complex line
parametrized by $c=\phi_+\phi_-$ to the single point $c=0$.  
Typically in the minimization of the potential of a SUSY gauge theory
one finds that the space of vacua (the ``moduli space'') may consist
of several sectors  
corresponding to ``flat directions'' running out of various minima 
that would be 
isolated if a suitably smaller set of chiral multiplets had been used.

\section{Renormalizable Model with Triplets}

Let us now consider the pattern of symmetry breaking at $M_R >> M_S\sim M_W$.
We assume that the soft terms are such as to keep squark and bi-doublet vevs
zero since these would break colour, charge or the $SU(2)_L$
symmetry at the high scale. On the other hand since the neutral
component of the $L_c$ field could {\it{a priori }} get a v.e.v
without violating any low energy symmetry and we would like to
consider the case where L-R symmetry is spontaneously violated ,
we keep both the $L_c$ and the $L$ fields besides the triplets.
The most general renormalizable superpotential is then :

\begin{eqnarray}
 W_{LR}&=& {\bf h}_l^{(i)} L^T \tau_2 \Phi_i \tau_2 L_c 
       +i {\bf f} (L^T \tau_2 \Delta L +
 L^{cT}\tau_2 \Delta_c L_c) \nonumber \\
& & +  m_\Delta ( {\rm  Tr}\, \Delta \overline{\Delta}    
  +   {\rm Tr}\,\Delta_c \overline{\Delta}_c ) 
+{m_{\Omega} \over 2} ( {\rm Tr}\,\Omega^2 +
           {\rm Tr}\,\Omega_c^2 )
\nonumber \\
& &  
 + \mu_{ij} {\rm Tr}\,  \tau_2 \Phi^T_i \tau_2 \Phi_j 
     +a ({\rm Tr}\,\Delta \Omega \overline{\Delta} +
        {\rm Tr}\,\Delta_c \Omega_c \overline{\Delta}_c )
\label{superpot}
\end{eqnarray}

Multiplying the the F flatness conditions for the triplet
 equations by triplet fields and taking traces it immediately follows

\begin{eqnarray}
&{\rm Tr}\,\Delta^2 = {\rm Tr}\,\Delta
\Omega = {\rm Tr}\,\overline\Delta
\Omega= 0\nonumber \\
& m_\Delta {\rm Tr}\,\Delta {\overline\Delta} =
m_\Omega  {\rm Tr}\, \Omega^2 =
a {\rm Tr}\,\Delta {\overline\Delta} \Omega 
\nonumber \\
&{\rm Tr} \Delta {\overline \Delta} \left(a^2 {\rm Tr }\Omega^2 - 2
m_\Delta^2 \right) = 0 
\label{fixers} 
\end{eqnarray}

with corresponding equations  {\it mutatis mutandis } in the right
handed sector. 
Thus it is clear that in either sector all three triplets are zero or non 
zero together. By 
choosing the branch where ${\rm Tr}\, \Omega_c^2 = 2 m_\Delta^2/
a^2 $ but ${\rm Tr}\, \Omega^2 =0$ we ensure that the triplet vevs break
$SU(2)_R$ but not $SU(2)_L$.

One can use the 3 parameters of the $SU(2)_R$ gauge freedom 
to set the 
diagonal elements of $\Delta_c$ to zero so that it takes the form 
\begin{equation}
\left < \Delta_c \right > =\left ( \matrix{
                      0 &\langle \delta_c^{--}\rangle \cr
                     \langle\delta_c^0\rangle &0 \cr } \right )
\label{rotation}
\end{equation}
Now (\ref{fixers}) gives $\langle\delta_c^{--}\rangle\langle 
\delta_c^0\rangle = 0$, which implies the electromagnetic
 charge-preserving form for  $\langle \Delta_c \rangle$.
Next it is clear that the Majorana coupling matrix $f_{ab}$ must be 
non-singular if the 
see saw mechanism \cite{seesaw} which keeps the neutrino light is to operate.
 Then it immediately follows from the condition $F_{L_c}=0$, namely, 
\begin{equation}
2 i {\bf f}_{ab}\left ( \matrix{
                      0 & 0 \cr
                     \langle\delta_c^0\rangle &0 \cr } \right )
\left ( \matrix{\nu_c \cr e_c \cr} \right )^b =0
\label{flexplicit}
\end{equation}
 that the sneutrino 
vevs in the 
right-handed sector must vanish. Thus any vev of $L_c$ that appears at the 
high scale 
must necessarily break charge. We ensure that it (together with L) 
vanishes by suitably 
positive soft masses, just as for the squarks. When the sleptons
have zero vevs one can show that all holomorphic invariants that
one can form from the triplets are fixed (at each of the allowed
minima) so the parity violating vacuum is isolated. and in fact
the vevs can be shown to necessarily take the form 
\begin{eqnarray}
 \langle \Omega_c \rangle  = \left (\matrix{ w & 0 \cr
                    0 & -w \cr} \right ), &\quad &
 \langle L_c  \rangle = 0 \nonumber \\
\langle \Delta_c \rangle = 
  \left (\matrix{  0 & 0 \cr
                     d & 0 \cr } \right ), &\quad &
\langle \overline\Delta_c \rangle  =  
  \left (\matrix{  0 & \overline d \cr
                     0 & 0 \cr }\right )
\label{vev1}
\end{eqnarray} 

In fact using the $B-L$ gauge invariance to fix the relative phase of
 $d$ and ${\overline d}$ one obtains

\begin{equation}
 w  =  - {m_\Delta \over a} 
\equiv - M_R ,\quad
d = {\overline d} =  \left({ 2m_\Delta m_\Omega \over
{a}^2}\right)^{1/2} \equiv M_{BL}
\label{vev2}
\end{equation}

Furthermore one can show that  the inclusion of slepton
fields  leads to 9 (complex) flat directions out of the
discrete minima of the superpotential with triplets alone and
these are parametrized in a gauge invariant way by ($a,b..$ are
family indices)

\begin{equation}
z_{[ab] d}= \sigma_{[ab]} ( Tr {\bar{\Delta'}}^{(c)}_{dd}
{\bar\Delta}^{(c)})^{1\over 2} \sim {\tilde\nu}_{[a}
{\tilde e}_{b]} {\tilde e}^{(c)}_d
\end{equation}

where the composite multiplets $\sigma_{ab}$ (singlet) and
${\bar\Delta}_{ab}'$ (triplet) are defined by 

\begin{equation}
L_a {\tilde L}_b =  \sigma_{ab}   +
 {\overline\Delta'_{ab}}
\label{Fierz} 
\end{equation}
and similarly in the right handed sector.

\section{R Parity}

We have shown that R parity breaking at the high scale $M_R$
necessarily implies breaking of electromagnetism so that one
must assume that the soft terms are such as to forbid such
breaking . Then {\it{the effective theory below $M_{R,B-L}$ is the
MSSM with R parity (and a heavy $\nu^{(c)}$ superfield)  hence
a global B, L invariance !}}. However,
one may worry that,  due to the running of couplings, the
sneutrinos may obtain vevs ; analogously to the Higgs in the MSSM
which develops a vev inspite of a positive soft mass at high
scales due to the effect of its large yukawa coupling to the top
quark on its effective mass \cite{topev}.
 In the present case however this is
unlikely. Firstly ${\tilde{\nu}^{(c)}}$ has a large mass and can
shift away from zero only due to a linear term in the potential.
 Such a linear term can develop via the trilinear soft terms once
$\tilde\nu$ acquires a vev but is not present otherwise. On the
other hand a vev for the left handed sneutrino implies the
presence of a goldstone boson (the ``doublet majoron ''
\cite{am82})  with an appreciable  coupling to the
the low energy gauge group currents and is forbidden by the
precise measurements of the Z width at LEP. One final
possibility that one must consider is that the vev for
$\tilde\nu$ leads to a vev for ${\tilde{\nu}^{(c)}}$ which in
turn provides enough explicit L violation to give the putative
majoron a mass greater than $M_Z/2$ . However it is easy to see
that 

\begin{equation}
\langle \nu^c\rangle \simeq {m_S M_W \langle\nu\rangle \over M_{BL}^2} \, .
\end{equation}
This would lead to effective R-parity and global lepton
number violating terms of the  form $m_\epsilon^2 LH$ where 
\begin{equation}
m_\epsilon^2 \simeq {m_S^2 M_W \langle \nu \rangle \over M_{BL}^2}
\end{equation}
Then the ``Majoron'' would get a mass squared of order
\begin{equation}
m_J^2 \simeq m_{\epsilon}^2 {m_S\over \langle \nu \rangle} 
\simeq {m_S^3 M_W\over M_{BL}^2}
\end{equation}
Thus in order that $m_J$ be large enough to evade the width
bound the scale $M_{BL}$ would have to be $O(m_S)$ : which
possibility we do not consider here as discussed earlier.

Thus the effects of running cannot change our earlier
conclusions regarding the effective theory at low energies.

\section{Mass spectrum and Seesaw Mechanism} 
  In the renormalizable model  $SU(2)_R$ 
is broken down to $U(1)_R$ at a large scale
 $M_R = m_\Delta/a$, by the vev of $\Omega_c$.
Later the vevs of $\Delta_c$, $\overline\Delta_c$ are turned on at  
$M_{BL} = \sqrt{2 m_\Delta m_\Omega} /a$, breaking
 $U(1)_R \times U(1)_{BL}$ 
to $U(1)_Y$. However, a third scale appears in the superpotential,
$m_\Omega = M_{BL}^2/M_R$.
Although most of the extra fields of the renormalizable LR SUSY model
get masses at $M_R$ or $M_{B-L}$ there are two notable exceptions.

Firstly a fine tuning is necessary in order to keep one pair of
Electroweak doublets (out of the 2 pairs contained in the 2
bidoublets) light to serve to break the Electroweak group at $M_W$ :
\begin{equation}
\mu_{11} \mu_{22} - (\mu_{12}^2 - \alpha_{12}^2 M_R^2 ) \simeq M_W^2
\end{equation}

What is striking however is that the spectrum contains a
complete $SU(2)_L$ triplet ($\Omega$) of scalars and fermions with masses
$\sim M_{B-L}^2/M_R$. If the two scales are well separated this
mass could near the lower limit $\sim M_W$ imposed by the
consistency of the analysis.

Finally the seesaw mechanism takes its canonical form and 
is free of ``  contamination '' \cite{ms81} namely the  Majorana mass for
the left handed neutrino that arises in non-SUSY theories since
$\Delta$ aquires a small vev due to terms of form
$\Delta\Phi^2\Delta_c$ in the potential which lead to a linear
term for $\Delta$ after symmetry breaking.

\section{Non-Renormalizable model}

I now briefly describe the other alternative namely allowing non
renormalizable terms in the superpotential while retaining the
minimal set of fields \cite{ams97}.  As before one ignores the squarks and
bidoublets while analyzing the potenial at $M_{B-L}=M_R$ so that
the relevant superpotential to next to renormalizable order is

\begin{eqnarray}
W_{nr} &=&  m ({\rm Tr}\, \Delta \bar \Delta +
  {\rm Tr}\, \Delta_c {\bar \Delta}_c) +
i {\bf f} ( L^T \tau_2 \Delta L+  L_c^T\tau_2 \Delta_c L_c )
\nonumber \\
& & + {a \over 2 M}\left[  ({\rm Tr}\, \Delta \bar\Delta)^2
 +  ({\rm Tr}\, \Delta_c {\bar\Delta}_c)^2 \right]+ {c \over M}{\rm Tr}\,
 \Delta \bar\Delta {\rm Tr}\, \Delta_c \bar\Delta_c \nonumber \\
& & + {b \over 2 M} \left[ {\rm Tr}\, \Delta^2 {\rm Tr}\, \bar \Delta^2 +
 {\rm Tr}\, \Delta_c^2 {\rm Tr}\, \bar \Delta_c^2 \right] 
+ {{\bf k}_{ll} \over M}L^T \tau_2 L\,L_c^T \tau_2 L_c +
 \nonumber \\
& & + {1 \over M} \left[d_1{\rm Tr}\, \Delta^2 {\rm Tr}\,  \Delta_c^2 +
d_2 {\rm Tr}\, \bar \Delta^2 {\rm Tr}\, \bar \Delta_c^2 \right] + \ldots
\label{nonsuperpot}
\end{eqnarray}
The nonrenormalizable terms are suppressed by inverse powers of
a large scale $M$ (for instance the Planck mass). 
If a renormalizable interaction (such as $(L_cL_c)^2$) is already
present then we can safely neglect corrections $\sim M^{-1}$.
Now using  arguments exactly similar to those we gave for the
renormalizable case one can show that the conjugate sneutrino
vev must vanish at the high scale where SUSY is good and hence
R parity and electric charge can only be broken together. Thus
one must again invoke soft terms to avoid this unpleasant
possibility. With $L,L_c$ vevs zero one uses the holomorphic
invariants technique to show that there is an isolated parity
violating vacuum (besides the trivial one with unbroken symmetry
and other charge violating ones ). The detailed analysis given in 
\cite{amrs97} is a good illustration of the power of this
technique. . The scale of symmetry breaking is the geometric
mean of the mass parameter $m/a$ and the large scale $M$. Thus
if $M\sim M_P$ then $m\geq 1 TeV$ gives $M_R \geq 10^{11} GeV$.

As before one can finetune to keep a pair of doublets light.
However the what is remarkable is that there is a plethora of
supermultiplets with masses as low as $m\sim 1 TeV$ namely
one neutral and two doubly charged fields from the triplets
$\Delta_c, {\bar\Delta}_c$ together with two complete $SU(2)_L$ triplets 
$\Delta, {\bar\Delta}$ and two $SU(2)_L$ doublets ({\it{ over and above 
 the pair of $SU(2)_L$ doublets of the MSSM}}). Thus such ``minimal''
Left-Right Supersymmetric model have\cite{ams97,amrs97}
 the spectacular signature
of 4 distinct light doubly charged supermultiplets. It appears\cite{cm97}
that this property is quite robust since it is connected with
the fact that without non-renormalizable terms the vacuum breaks
charge so that the addition of these relatively small
corrections can only lift the mass of the supermultiplet
associated with the charge breaking flat directions by a small
amount. Furthermore accurately measured quantities such as the Z
width are sensitive to the presence of such particles in the low
energy spectrum so that they actually constrain\cite{cm97} $M_R$ to lie
above  $10^9$ GeV in such theories.
The presence of 4 massless $SU(2)_L$ doublets down to scales $\sim $ 1 TeV
 has important implications for the solutions to the
strong CP problem based on parity \cite{mr96,mrs97}.

The seesaw mechanism in this case is also quite distinct since 
it is `` impure'' i.e the potential contains terms
of form 
\begin{equation}
{m \over M} \Phi^2 \overline\Delta_c \Delta + m^2 \Delta^2 
\end{equation}
which gives a vev for $\Delta$
\begin{equation}
\langle\Delta\rangle \simeq { \langle\Phi^2\rangle \over \sqrt{m M}} \sim
{M_W^2 \over M_R}
\end{equation}
exactly as in the non-supersymmetric case \cite{ms81}.

\section{Conclusions} 

We have shown that Left Right Supersymmetric theories offer an
attractive matrix in which the MSSM can be embedded with a gain
of rationality as far as its problems with naturality and
options vis vis neutrino masses are concerned. The theory makes
definite testable predictions regarding the low energy spectrum
in both the viable and consistent ``minimal '' models analyzed.
In particular the non-renormalizable minimal models predict low
mass doubly charged supermultiplets that place\cite{cm97} lower bounds on
the scale $M_R$.
We have made extensive use of the powerful technique of holomorphic
invariants while analyzing the potential and its possible flat
directions. This permitted a clear and intelligible picture of
the low energy theory to emerge inspite of the plethora of
multiplets involved. Our work should provide a fresh impetus to
analysis of GUTS with a  left right symmetric intermediate
symmetry and serve as reference point for future work on such
theories.

\end{document}